%% file: main.tex
\documentclass[preprint,12pt]{elsarticle}

\usepackage{amssymb}
\usepackage{amsmath}
\usepackage{graphicx}
\usepackage{here}
\usepackage{mathrsfs}
\usepackage{bbm}
\usepackage{physics}
\usepackage{bm}
\usepackage{comment}
\usepackage{mhchem}
\usepackage{siunitx}
\usepackage{url}
\usepackage{subcaption}

\journal{Computer Physics Communications}

\begin{document}

\begin{frontmatter}



%\title{Symmetry-aware construction of the first-principles ferromagnetic Wannier model \\
%and its application to magnetocrystalline anisotropy}
\title{Efficient calculation of magnetocrystalline anisotropy energy 
using symmetry-adapted Wannier functions}


\author{Hiroto Saito}

\affiliation{organization={Department of Physics, Tohoku University},%Department and Organization
            addressline={Aoba-ku}, 
            city={Sendai},
            postcode={980-8578}, 
            % state={State One},
            country={Japan}}

\author{Takashi Koretsune}
% \author[inst1,inst2]{Author Three}

% \affiliation[inst2]{organization={Department Two},%Department and Organization
%            addressline={Address Two}, 
%            city={City Two},
%            postcode={22222}, 
%            state={State Two},
%            country={Country Two}}

\input{./abstract.tex}


% \begin{highlights}
% \end{highlights}


\end{frontmatter}


\input{./introduction.tex}
\input{./method.tex}

\input{./result.tex}
\input{./result1.tex}
\input{./conclusion.tex}
\input{./appendix.tex}

 \bibliographystyle{elsarticle-num} 
 \bibliography{main}





% \end{thebibliography}
\end{document}

%% file: abstract.tex
%! Author = Hiroto Saito
%! Date = 2023/12/22

\begin{abstract}
    Magnetocrystalline anisotropy, a crucial factor in magnetic properties and applications like magnetoresistive random-access memory, often requires extensive $k$-point mesh in first-principles calculations. 
    In this study, we develop a Wannier orbital tight-binding model incorporating crystal and spin symmetries 
    and utilize time-reversal symmetry to divide magnetization components. 
    This model enables efficient computation of magnetocrystalline anisotropy. 
    Applying this method to $\mathrm{L1_0}$ $\mathrm{FePt}$ and $\mathrm{FeNi}$, 
    we calculate the dependence of the anisotropic energy on $k$-point mesh size, chemical potential, spin-orbit interaction, and magnetization direction.
    The results validate the practicality of the models to the energy order of $10~[\mathrm{\mu eV}/f.u.]$.
\end{abstract}

%% file: introduction.tex
%! Author = Hiroto Saito
%! Date = 2023/12/22

\section{Introduction}
The Maximally localized Wannier functions (MLWF) method is a widely utilized post-processing approach in first-principles calculations \cite{Marzari1997-rw,Souza2001-hk,Marzari2012-xr,Pizzi2020-ac}. 
It is primarily used for generating localized basis sets from band structures, 
which are crucial in finely interpolating the $k$-mesh of the energy bands and in constructing a model Hamiltonian.
The MLWF method is distinguished from other interpolation methods by its requirement of 
not only the energy eigenvalues at each $k$-point but also the connectivity of eigenstates. 
This aspect makes it particularly effective in accurately reproducing first-principles bands, 
even in cases involving band crossings or degeneracies \cite{Yates2007-jf}.

Wannier functions (WFs) are usually constructed through a process that involves iteratively minimizing the sum of the spreads. 
This is achieved by employing 
the algorithm of Marzari and Vanderbilt for an isolated group of bands \cite{Marzari1997-rw} and the disentanglement scheme for entangled bands \cite{Souza2001-hk}.
However, these processes do not ensure that the resulting WFs will follow the spatial symmetry of the crystal. 
Preserving this symmetry is critical for the physically interpretable model Hamiltonian based on the results of first-principles calculations. 
To address this issue, the symmetry-adapted Wannier function (SAWF) method was developed \cite{Sakuma2013-rq}. 
This method generates the SAWFs by applying additional constraints based on the operations of the site symmetry group during the maximally localization process. 
Recently, this method has been further extended to the disentanglement process with a frozen (inner) window \cite{Koretsune2023-ev}.

In general, when calculating physical properties in metals, 
the convergence speed of integrations of the Brillouin zone (BZ) is slower compared to the case of insulators, 
due to partially filled energy bands. 
This necessitates a significantly large number of $k$-points, posing a challenge to the efficiency and accuracy of first-principles calculations for many physical properties.
One notable example of such challenges is the calculation of the magnetocrystalline anisotropy (MCA) in ferromagnetic materials. 
The MCA refers to the phenomenon where the internal energy of a material varies with the direction of magnetization, and it is a crucial property of magnetic materials \cite{Miura2022-ta}. 
In practical applications, materials used in magnetoresistive random-access memory (MRAM) require perpendicular anisotropy to enhance the thermal stability of the magnetization direction \cite{Bhatti2017-xi}. 
The importance of in-plane anisotropy is also emphasized in applications such as 
soft magnetic underlayers for perpendicular recording media \cite{Hashimoto2006-ad} and spin-torque oscillators for microwave-assisted magnetic recording \cite{Yoshida2010-ev,Igarashi2010-ke}.

To elucidate the microscopic origins of the MCA, extensive research has been conducted across theoretical, computational, and experimental fields. 
Theoretically, the primary mechanisms of the MCA in transition metal compounds have been explained through the second-order perturbation theory of spin-orbit interaction (SOI), 
as proposed by Bruno \cite{Bruno1989-lc} and extended by Van der Laan \cite{Van_der_Laan1998-rm}. 
In computational studies, first-principles calculations have been vigorously pursued, 
particularly for compounds with $\mathrm{L1_0}$ structures 
\cite{Daalderop1991-ul,Sakuma1994-em,Solovyev1995-lt,Oppeneer1998-fb,Galanakis2000-jm,Ravindran2001-ts,Staunton2004-wn,Burkert2005-lu,Lu2010-ee,Kosugi2014-zl,Ayaz_Khan2016-qr,Ke2019-lm,Alsaad2020-jf,
Miura2013-nl,Edstrom2014-ok,Qiao2019-nc,Woodgate2023}. 
Notably, Khan $et~al.$ \cite{Ayaz_Khan2016-qr} have investigated higher-order SOI effects by self-consistently solving the Dirac equation, 
examining the impact of different exchange-correlation functionals on many-body effects, 
and evaluating the validity of the magnetic force theorem for $\ce{FePt}$.
More recent efforts include detailed studies of the convergence of the MCA energy (MCAE) to the number of $k$ points using the Wannier interpolation method \cite{Qiao2019-nc}, 
as well as calculations of the temperature dependence of the MCAE using the disordered local moment method, a form of the coherent potential approximation \cite{Yamashita2022-sp,Woodgate2023}. 
On the experimental front, the MCAE is typically determined by measuring the magnetic hysteresis curves in two directions. 
Additionally, by measuring the torque dependent on the direction of magnetization, higher-order magnetic anisotropy constants have been determined 
\cite{Okamoto2002-np,Inoue2006-cp,Richter2011-uo,Ono2018-pp,Neel1964-el,Pauleve1968-qo}. 
Despite these experimental findings, computational calculations of the angular dependence of the MCAE, particularly the subtle anisotropies within the easy plane, have been limited.

In this study, we first calculate the MCAE of $\mathrm{L1_0}$-structured $\ce{FePt}$ and $\ce{FeNi}$ 
by performing $k$-point interpolation using a model Hamiltonian based on the SAWF method. 
Furthermore, by exploiting time-reversal symmetry operation in the model Hamiltonian, 
we extract and rotate the magnetization to determine the dependence of the MCAE on the full angular range of magnetization. 
This method is computationally cost-effective, as it allows for the generation of a model Hamiltonian with magnetization oriented in any direction through a single Wannierization process.

%% file: method.tex
%! Author = Hiroto Saito
%! Date = 2023/12/22

\section{Method}
First, we summarize the SAWF method developed in Refs.\ \cite{Sakuma2013-rq} and \cite{Koretsune2023-ev}.
We then describe the symmetry of the spin space considered.
Next, we discuss the model Hamiltonian and the approximations.
Finally, we give the specific calculation method of the MCAE.

\subsection{Symmetry-adapted Wannier functions}
We denote the space group of the system as \(\mathcal{G}\) and its element as \(g \in \mathcal{G}\). 
Within the space group \(\mathcal{G}\), we identify a subgroup of elements that leave a certain wave vector \(\bm{k}\) unchanged. 
This subgroup is referred to as the little group of \(\bm{k}\), denoted as \(G_{\bm{k}}\), 
\begin{align}
    G_{\bm{k}} = 
    \{h \in \mathcal{G} ~|~ h\bm{k} \doteq \bm{k} \} .
\end{align}
Here, the symbol \(\doteq\) signifies equality allowing for the difference in reciprocal lattice vectors.

Generally, WFs are defined through Fourier transformation combined with projections of Bloch wave functions and their unitary transformation,   
\begin{align}
    &\ket{\psi^{\mathrm{opt}}_{m\bm{k}}}
    = \sum_{l=1}^{N_{\mathrm{band}}}
    \ket{\psi_{l\bm{k}}} U^{\mathrm{opt}}_{lm,\bm{k}} , \label{eq:disentanglement} \\
    &\ket{W_{n\bm{R}}}
    = \frac{V}{(2\pi)^3} \int \dd[3]{k} e^{-i\bm{k}\cdot\bm{R}}
    \sum_{m=1}^{N_{\mathrm{wann}}} \ket{\psi^{\mathrm{opt}}_{m\bm{k}}} U_{mn,\bm{k}} . \label{eq:fourier}
\end{align}
Eq.\ (\ref{eq:disentanglement}) represents the process of disentanglement, where \(\ket{\psi^{\mathrm{opt}}_{m\bm{k}}}\) is an optimized subspace derived from a larger set of Bloch bands \(\ket{\psi_{l\bm{k}}}\). 
The matrix \(U^{\mathrm{opt}}_{\bm{k}}\) is a projection matrix of dimensions \(N_{\mathrm{band}} \times N_{\mathrm{wann}}\), 
where \(N_{\mathrm{band}}\) is the number of Bloch bands considered, and \(N_{\mathrm{wann}}\) is the number of WFs being projected, satisfying \(N_{\mathrm{wann}} \leq N_{\mathrm{band}}\).
Eq.\ (\ref{eq:fourier}) represents the Fourier transformation, 
where \(\bm{R}\), \(V\), and \(U_{\bm{k}}\) represent a Bravais lattice vector, the volume of the unit cell, and a \(N_{\mathrm{wann}} \times N_{\mathrm{wann}}\) unitary matrix, respectively.
The total wave-vector-dependent gauge \(U^{\mathrm{tot}}_{ln,\bm{k}}\) is defined as
\begin{align}
    U^{\mathrm{tot}}_{ln,\bm{k}} = \sum_{m=1}^{N_{\mathrm{wann}}} U^{\mathrm{opt}}_{lm,\bm{k}} U_{mn,\bm{k}} .
\end{align}

To apply the symmetry constraints for the gauge \(U^{\mathrm{tot}}_{\bm{k}}\), there are two steps. 
The first one is to ensure that \(U^{\mathrm{tot}}_{\bm{k}}\) for $k$ points in the irreducible BZ (IBZ) remains invariant under the operations of the little group \(G_{\bm{k}}\).
This step involves replacing \(U^{\mathrm{tot}}_{\bm{k}}\) with \(\overline{U}^{\mathrm{tot}}_{\bm{k}}\), which averages the transformations corresponding to all elements of the little group,  
\begin{align}
    \overline{U}^{\mathrm{tot}}_{\bm{k}} 
    = \frac{1}{|G_{\bm{k}}|} \sum_{h\in G_{\bm{k}}} h U^{\mathrm{tot}}_{\bm{k}} h^{-1} .
    ~~~~(\bm{k} \in \mathrm{IBZ}) \label{eq:ave}
\end{align}
Next, the IBZ is expanded to encompass the full BZ by using symmetry operations \( g \in \mathcal{G} \) of the space group, 
\begin{align}
    \overline{U}^{\mathrm{tot}}_{g\bm{k}} 
    = g \overline{U}^{\mathrm{tot}}_{\bm{k}} g^{-1}.
    ~~~~(\bm{k} \in \mathrm{IBZ})  \label{eq:IBZ to BZ}
\end{align}
Here, the representation matrices of operators, $g$ and $h$, are determined by the Bloch wave functions and Wannier functions. By applying this symmetry constraint, we can obtain the SAWFs.
Ref.\ \cite{Sakuma2013-rq} considered the symmetrization of both \( U^{\mathrm{opt}}_{\bm{k}} \) and \( U_{\bm{k}} \) according to Eqs.\ (\ref{eq:ave}) and (\ref{eq:IBZ to BZ}). 
In contrast, Ref.\ \cite{Koretsune2023-ev} applied the symmetrization only to \( U^{\mathrm{tot}}_{\bm{k}} \).

\subsection{Time-reversal symmetric Wannier functions}
When constructing the tight-binding model for ferromagnetic materials, one sometimes considers the magnetic moment along the $z$ direction and constructs the Wannier functions for $S_z=\pm 1/2$. 
In this approach, Wannier functions for $S_z=1/2$ and $S_z=-1/2$ are different. 
This can be understood using the site symmetry group that there is no site symmetry operation that connects a Wannier function with $S_z = 1/2$ to that with $S_z = -1/2$.
On the other hand, when constructing the Wannier functions for $S_x=\pm 1/2$, two Wannier functions can have same wave functions for the real space. 
In other words, time-reversal symmetric WFs (TRS-WFs) can be obtained. 
This can be understood that site symmetry group can have a symmetry operation that connects Wannier functions for $S_x = 1/2$ and that for $S_x = -1/2$. 
Of course, whether such symmetry exists depends on the crystal symmetry, while many crystal structures have such symmetry operations. 
Fig.\ 1 schematically illustrates the construction of TRS-WFs.

\begin{comment}
Furthermore, this study constructs time-reversal symmetric WFs (TRS-WFs), whose centers and spreads are independent of spin, 
by implementing a wave vector-independent gauge fixing in the internal spinor space. 
This addition to the symmetry conditions of the space group is depicted in figure \ref{fig:TRS Wannier}. 
The feasibility of consistently creating such TRS-WFs, from the perspective of topological symmetry indicators, is not a trivial issue. 
However, as demonstrated in section \ref{sec:calculation details}, this condition has been numerically verified in the present research.
\end{comment}

\begin{figure}[H]
    \centering
    \includegraphics[scale=0.3]{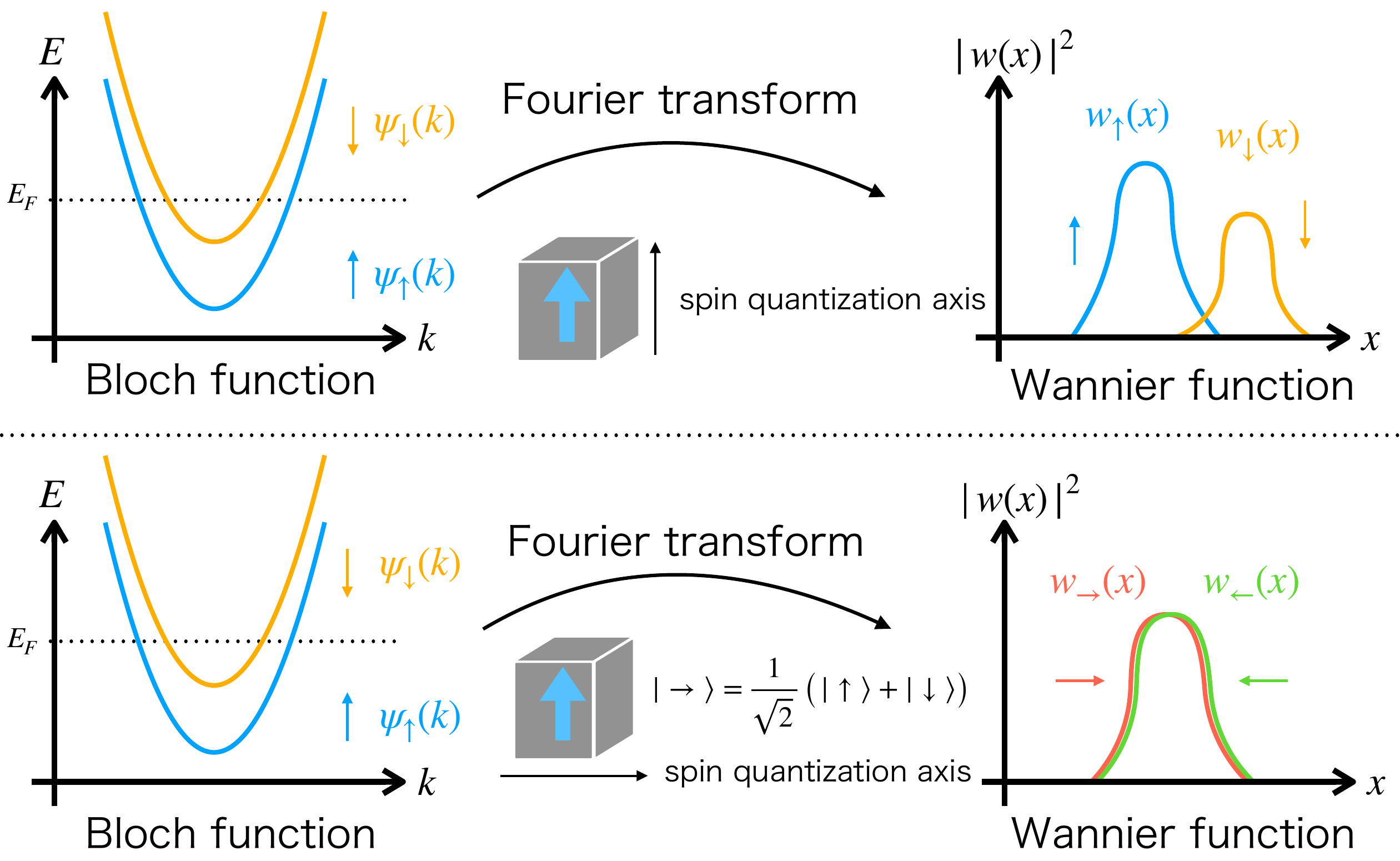}
    \caption{Schematic diagram of the process of constructing time-reversal symmetric Wannier functions. 
    At the top, typical Wannier functions for up spin and down spin in a ferromagnetic material are depicted. 
    In cases where the spin quantization axis for Wannierization aligns with the direction of magnetization, 
    the spatial distributions of these Wannier functions do not match.
    At the bottom, time-reversal symmetric Wannier functions are shown. 
    By orthogonalizing the spin quantization axis for Wannierization to the direction of magnetization, 
    Wannier functions are obtained where the centers and spreads of the functions are paired, thereby matching. }
    \label{fig:TRS Wannier}
\end{figure}

\subsection{Model Hamiltonian}
\label{sec:model hamiltonian}
The model Hamiltonian, dependent on the direction of magnetization, was derived following these steps:
Initially, under the basis of TRS-WFs \(\ket{j\mu\sigma}\), which are localized at position \(j\) and specified by orbital \(\mu\) and spin index \(\sigma\), 
the matrix elements of the tight-binding Hamiltonian were decomposed into two terms: one symmetric and the other antisymmetric with respect to time reversal, 
\begin{align}
    \mel{j\mu\sigma}{H}{k\nu\sigma'} 
    = H_{j\mu\sigma; k\nu\sigma'}
    = H_{j\mu\sigma; k\nu\sigma'}^s 
    + H_{j\mu\sigma; k\nu\sigma'}^a  .
    \label{eq:TRS ham}
\end{align}
Here, the time-reversal operator, 
\begin{align}
    \Theta = -i\sigma^y K 
    = 
    \begin{pmatrix}
        0 & -1 \\
        1 & 0
    \end{pmatrix} K \label{eq:TR op}
\end{align}
is utilized,
where \(\sigma^y\) is the \(y\) component of the Pauli matrices \(\bm{\sigma} = (\sigma^x, \sigma^y, \sigma^z)\) 
and the operator \(K\) represents the complex conjugation.

Subsequently, the first term of Eq.\ (\ref{eq:TRS ham}) was further divided into two parts: 
one independent of spin, identified as the hopping integral \(t\), and the other dependent on spin, 
recognized as the SOI represented by \(\bm{\lambda} = (\lambda^x, \lambda^y, \lambda^z)\) \cite{Kurita2020-mh}, 
\begin{align}
    H_{j\mu\sigma; k\nu\sigma'}^s 
    = t_{j\mu; k\nu} \delta_{\sigma\sigma'}
    + i( \bm{\lambda}_{j\mu; k\nu} \cdot \bm{\sigma})_{\sigma\sigma'} .
    \label{eq:ham sym}
\end{align}
It is easy to show that due to the properties of the time-reversal operator \(\Theta\), 
all components of the hopping integral \(t\) and the SOI \(\bm{\lambda}\) are fundamentally real numbers.
Similarly, the second term of Eq.\ (\ref{eq:TRS ham}) can be decomposed using Pauli matrices,  
\begin{align}
    H_{j\mu\sigma; k\nu\sigma'}^a
    = i M^0_{j\mu; k\nu} \delta_{\sigma\sigma'}
    + ( \bm{M}_{j\mu; k\nu} \cdot \bm{\sigma})_{\sigma\sigma'} .
    \label{eq:ham asym}
\end{align}
The components \(M^0\) and \(\bm{M} = (M^x, M^y, M^z)\) of this term are also all real numbers.

In this study, we calculated the MCAE using three distinct methods.
Method 1: The MCAE is calculated by rotating the magnetization in the tight-binding model defined in Eq.\ (\ref{eq:ham asym}). 
In this approach, $H^a$ is approximated as the principal component of magnetization, \(M^z_{i\mu; j\nu}\), 
\begin{align}
    H_{j\mu\sigma; k\nu\sigma'}^a
    \approx M^z_{i\mu; j\nu} (\sigma^z)_{\sigma\sigma'}, \label{eq:effective mag}
\end{align}
and rotated with the usual spinor rotation, 
\begin{align}
    U_\mathrm{s} (\theta,\phi) = 
\begin{pmatrix} 
    \cos\theta/2 & e^{-i\phi} \sin\theta/2 \\
    -e^{i\phi} \sin\theta/2 & \cos\theta/2 \label{eq: spinor rotation}
\end{pmatrix}. 
\end{align}
Indeed, when performing fully relativistic calculations, the assumption of collinear magnetization as the self-consistent spin density is justified up to the fourth-order perturbations of the SOI, according to the magnetic force theorem \cite{Weinert1985-ib,Wang1996-ba}.
We call this rotated tight-binding Hamiltonian as the tight-binding model with effective magnetization (TB-EM).
The average value of the magnetization matrix elements discarded in the TB-EM method is defined as the computational error, 
\begin{align}
    E_{\mathrm{err}} 
    = \frac{1}{|H|}
    \sum_{jk} \sum_{\mu\nu} \sum_{\sigma\sigma'}
    \{H_{j\mu\sigma; k\nu\sigma'}^a - (M^z_{i\mu; j\nu}\sigma^z)_{\sigma\sigma'} \}
\end{align}
where \(|H|\) represents the number of matrix elements in the Hamiltonian. 
This computational error, \(E_{\mathrm{err}} \), %arises from the modulation of the ideally real WFs by the SOI and 
corresponds to terms higher than the fourth-order perturbation of the SOI.

Method 2: Similar to Method 1, the MCAE is calculated by rotating the magnetization in the tight-binding model. 
However, this method retains all terms in Eq.\ (9) for the spinor rotation, and thus, we call this method as the tight-binding model with original magnetization (TB-OM) method.

Method 3: Here, the MCAE is calculated by rotating the magnetization in density functional theory (DFT) calculations. 
The tight-binding models are constructed for each of the DFT calculations and are utilized only for interpolating the energy of bands.
We call this method as the DFT method.
In the results section, we compare these three methods to discuss their validity and effectiveness.

\subsection{Magnetocrystalline anisotropy energy}
The MCAE, denoted as \(\Delta E_{\mathrm{MCA}}\), is calculated based on the magnetic force theorem \cite{Weinert1985-ib,Wang1996-ba},
derived from the difference in band energies, 
\begin{align}
    \Delta E_{\mathrm{MCA}} (\theta, \phi)
    = \frac{1}{N_{k}} \sum_{\bm{k}}
     \left(
        \sum_{n} f(\varepsilon_{n\bm{k}}^{(\theta,\phi)}) \varepsilon_{n\bm{k}}^{(\theta,\phi)}
        - \sum_{m} f(\varepsilon_{m\bm{k}}^{(0,0)}) \varepsilon_{m\bm{k}}^{(0,0)}
    \right) .
\end{align}
Here, \(n\) and \(m\) are indices designating the bands, while \((\theta,\phi)\) represent the angles specifying the direction of magnetization.
The number of \(k\) points within the BZ is denoted by \(N_k\). 
The terms \(\varepsilon_{n\bm{k}}\) and \(f(\varepsilon_{n\bm{k}})\) represent the energy at band \(n\) and wave vector \(\bm{k}\), and the Fermi distribution function respectively, 
where \(f(\varepsilon) = \frac{1}{e^{\beta(\varepsilon - \mu)} + 1}\).

The Fermi energy \(\mu^{(\theta,\phi)}\), which depends on the direction of magnetization, 
is determined so as to satisfy the condition that the number of valence electrons per unit cell, \(N_{\mathrm{elec}}\), remains constant at the temperature $T=0$, 
\begin{align}
    N_{\mathrm{elec}} = \frac{1}{N_k} \sum_{\bm{k}} \sum_n f(\varepsilon_{n\bm{k}}^{(\theta,\phi)}) .
\end{align}

\subsection{Calculation details}
\label{sec:calculation details}
\begin{figure}[H]
    \centering
    \includegraphics[scale=0.15]
    {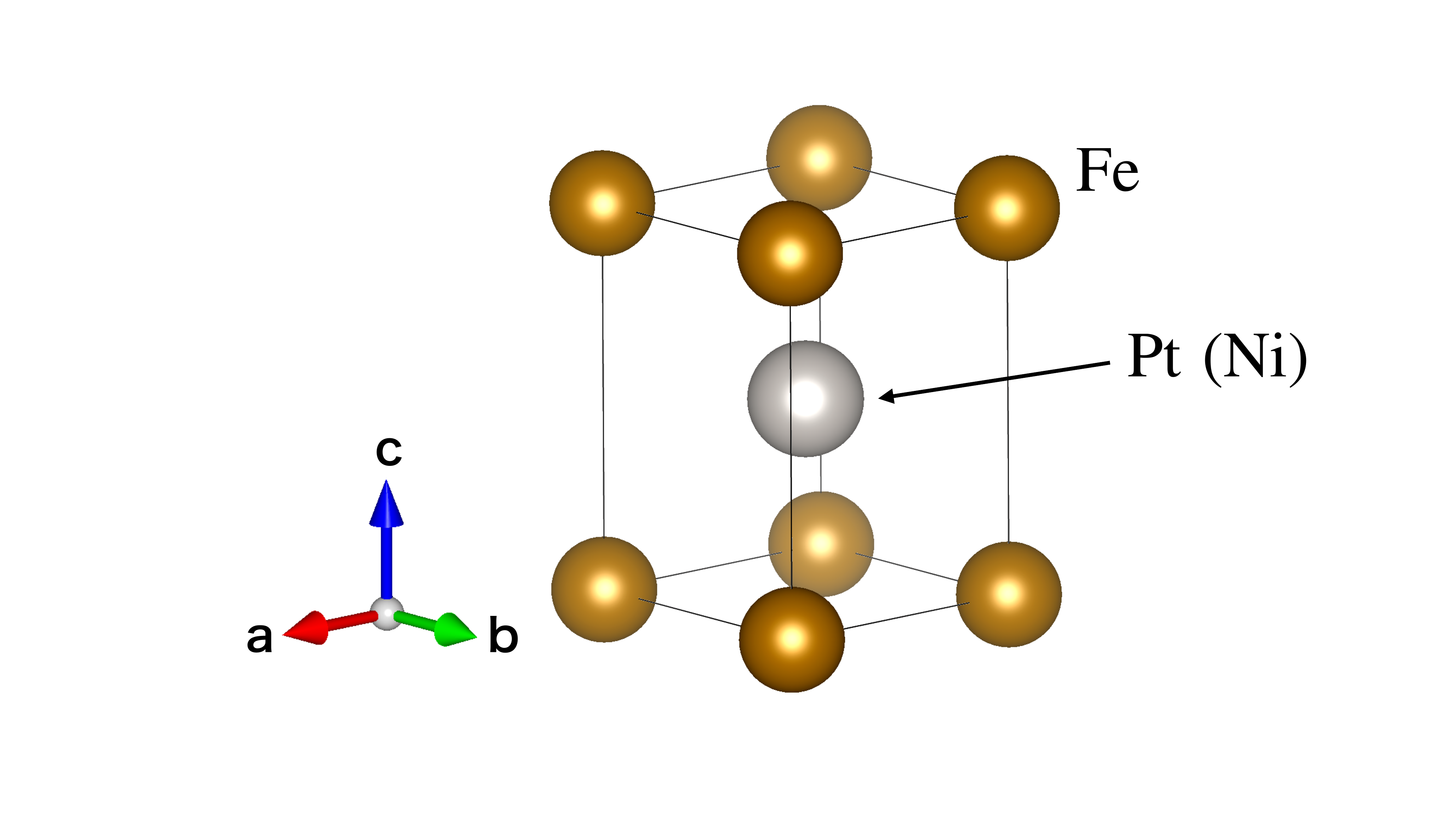}
    \caption{\(\mathrm{L1_0}\) type crystal structure, 
    where orange and grey spheres represent \(\ce{Fe}\) and \(\ce{Pt}\) (\(\ce{Ni}\)) atoms, respectively. 
    }
    \label{fig:crystal structure}
\end{figure}
First-principles calculations program {\sc{Quantum ESPRESSO}} \cite{Giannozzi2009-zd,Giannozzi2017-yu} based on plane-wave basis and pseudopotential methods were employed to obtain the Bloch states. 
Subsequently, a tight-binding model was created using {\sc{Wannier90}} program \cite{Pizzi2020-ac} in combination with {\sc{SymWannier}} program \cite{Koretsune2023-ev}, which takes into account the magnetic space group of the crystal.
For the \(\mathrm{L1_0}\) type crystal structure (Fig.\ \ref{fig:crystal structure}), lattice constants were set as 
\(a=2.728143~ [\si{\angstrom}]\), \(c= 3.778984~ [\si{\angstrom}]\) for \(\ce{FePt}\),
\(a=2.506611~ [\si{\angstrom}]\), \(c= 3.581316~ [\si{\angstrom}]\) for \(\ce{FeNi}\). 
The cutoff for the plane wave basis was set as \(64.0 ~[\mathrm{Ry}]\), 
and for the electron density as \(782.0 ~[\mathrm{Ry}]\). 
The exchange-correlation functional employed was the Perdew-Burke-Ernzerhof (PBE) formulation \cite{Perdew1996-tj} of the generalized gradient approximation (GGA).
Ultrasoft pseudopotentials from the {\sc{PSLibrary}} \cite{Dal_Corso2014-hd} were utilized, with version 0.2.1 for \(\ce{Fe}\) and version 0.1 for \(\ce{Pt}\) and \(\ce{Ni}\). 
According to the magnetic force theorem, SOI was incorporated during the non self-consistent field (NSCF) calculations.

Wannierization was carried out in a one-shot process, where the projection onto localized $s, p, d$ orbitals at the atomic center positions for each atom yielded 36 spinor Wannier Functions (WFs). 
For the $k$-mesh during Wannierization, a grid of $9 \times 9 \times 6$ was chosen for \(\ce{FePt}\) and $9 \times 9 \times 7$ for \(\ce{FeNi}\). 
From these, only $60$ inequivalent $k$ points under symmetry operations were utilized in the actual calculations for both cases.
The inner energy window for disentanglement was carefully set to include the all $3d$ states of the magnetic element \(\ce{Fe}\). 
For \(\ce{FePt}\), a range of \([-1000, E_F + 4] ~[\si{eV}]\) was used, and for \(\ce{FeNi}\), the range of \([-1000, E_F + 5] ~[\si{eV}]\) was used.
The comparison between DFT and Wannier interpolated band structure is explained in \ref{app:band and dos}.
Importantly, during Wannierization, the spin quantization axis (the $a$ axis) was set orthogonal to the direction of magnetization (the $c$ axis) in the first-principles calculations. 
This ensured that for both up and down spinors of the WFs, the differences in the centers of the WFs were less than \(10^{-6}~[\si{\angstrom}]\), 
and the differences in the spreads of the WFs were less than \(10^{-8}~[\si{\angstrom}^2]\) for all pairs.

%% file: result.tex
%! Author = Hiroto Saito
%! Date = 2023/12/22

\section{Result}
We first utilize the TB-EM method to increase the number of \(k\)-points in the BZ, 
to ensure the convergence of the MCAE.
Following this, the dependence of MCAE on the chemical potential and SOI is examined. 
This involves comparing the three methods described in section \ref{sec:model hamiltonian}, 
with a particular focus on assessing the validity of the TB-EM approach.
Lastly, we compute the dependence of the MCAE on the angle of magnetization to derive the magnetic anisotropy constants.

\subsection{$k$-mesh dependence}
\begin{figure}[H]
    \captionsetup[sub]{justification=justified,singlelinecheck=false} % subcaptionの位置を左に設定
    \begin{minipage}[b]{0.48\columnwidth}
    \centering
    \subcaption{}
    \includegraphics[scale=0.43]
    {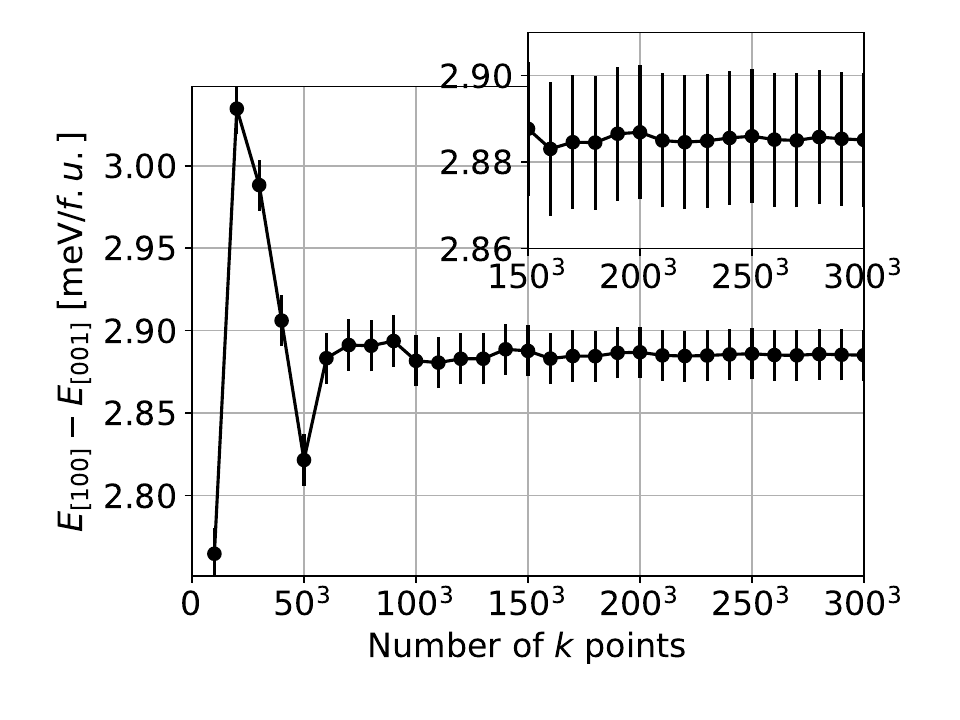}
    \end{minipage}
    \begin{minipage}[b]{0.48\columnwidth}
    \centering
    \subcaption{}
    \includegraphics[scale=0.43]
    {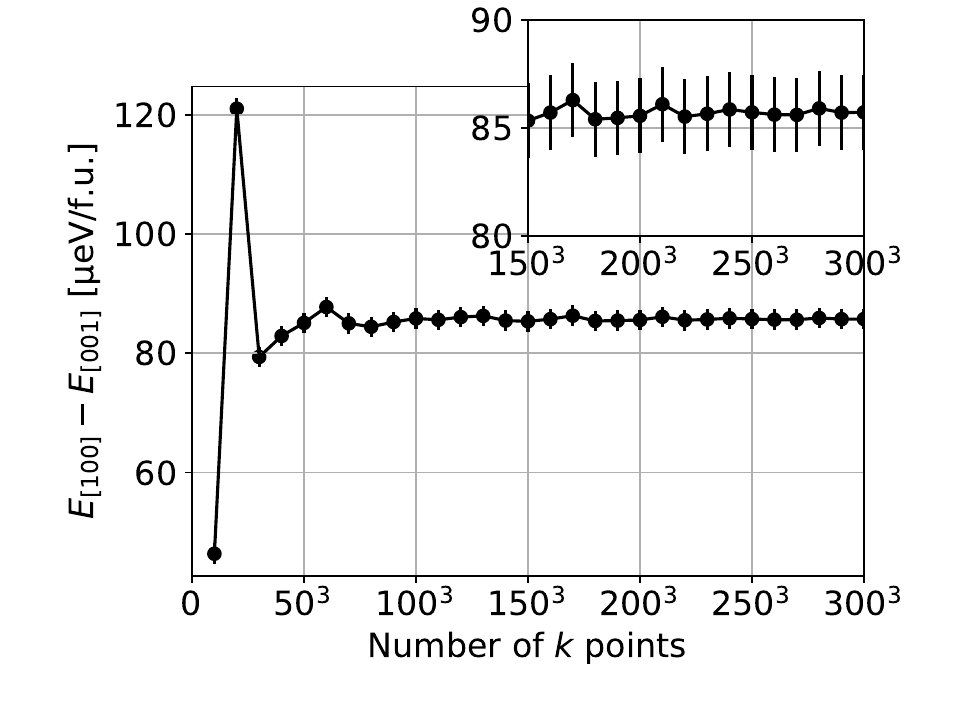}
    \end{minipage}
    \caption{
        MCAE for (a) \(\ce{FePt}\) and (b) \(\ce{FeNi}\) with respect to the number of $k$-points.
        The error bars indicate the respective computational errors for each calculation.
        The MCAE converges to 
        \(2.885 \pm 0.012~[\mathrm{meV}/f.u.] = 17.21 \pm 0.07~[\mathrm{MJ/m^3}]\) for FePt 
        and \(85.63 \pm 2.71~[\mathrm{\mu eV}/f.u.] = 0.610 \pm 0.019~[\mathrm{MJ/m^3}]\)  for FeNi, respectively.}
    \label{fig:mae_conv}
\end{figure}

We initially employ \(k\) interpolation with the TB-EM method to investigate the degree of convergence of the MCAE 
between the hard axis \([100]\) and the easy axis \([001]\), represented as \(E_{[100]} - E_{[001]}\).
The results are presented in Fig.\ \ref{fig:mae_conv}. 
It is observed that for both \(\ce{FePt}\) and \(\ce{FeNi}\), increasing the number of \(k\)-points to \(N_k=100^3\) ensures sufficient convergence of the MCAE 
within the range of computational error $E_{\mathrm{err}}$. 
In all subsequent calculations presented in this paper, we used a uniform \(k\)-point grid of \(N_k=100^3\) points to determine the other dependence.

\subsection{Chemical potential dependence}
\begin{figure}[H]
    \captionsetup[sub]{justification=justified,singlelinecheck=false} % subcaptionの位置を左に設定
    \begin{minipage}[b]{0.48\columnwidth}
    \centering
    \subcaption{}
    \includegraphics[scale=0.43]
    {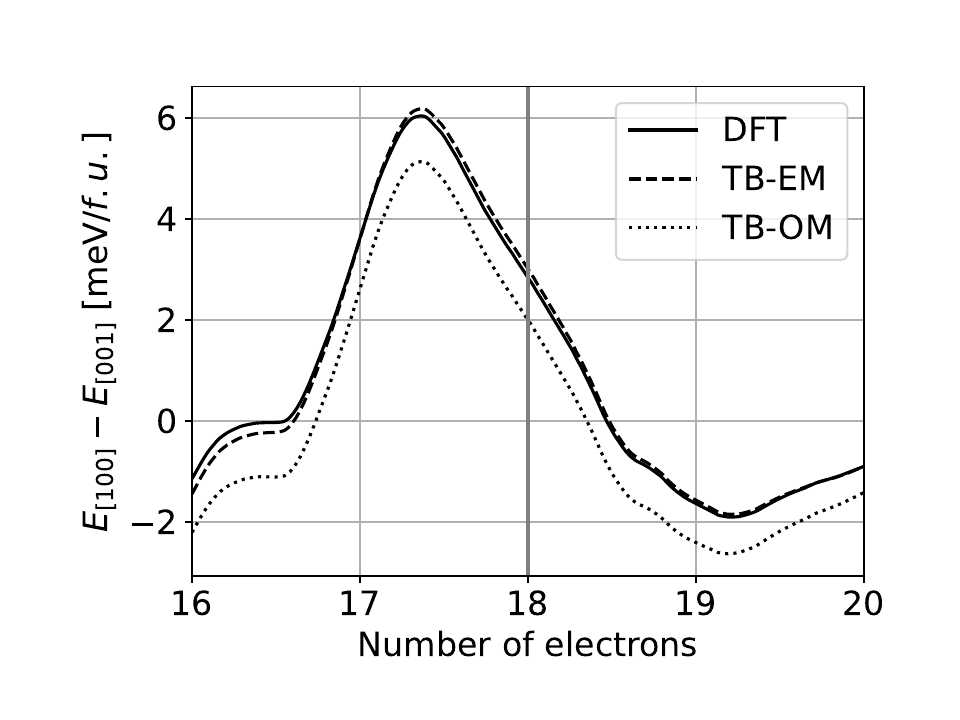}
    \end{minipage}
    \begin{minipage}[b]{0.48\columnwidth}
    \centering
    \subcaption{}
    \includegraphics[scale=0.43]
    {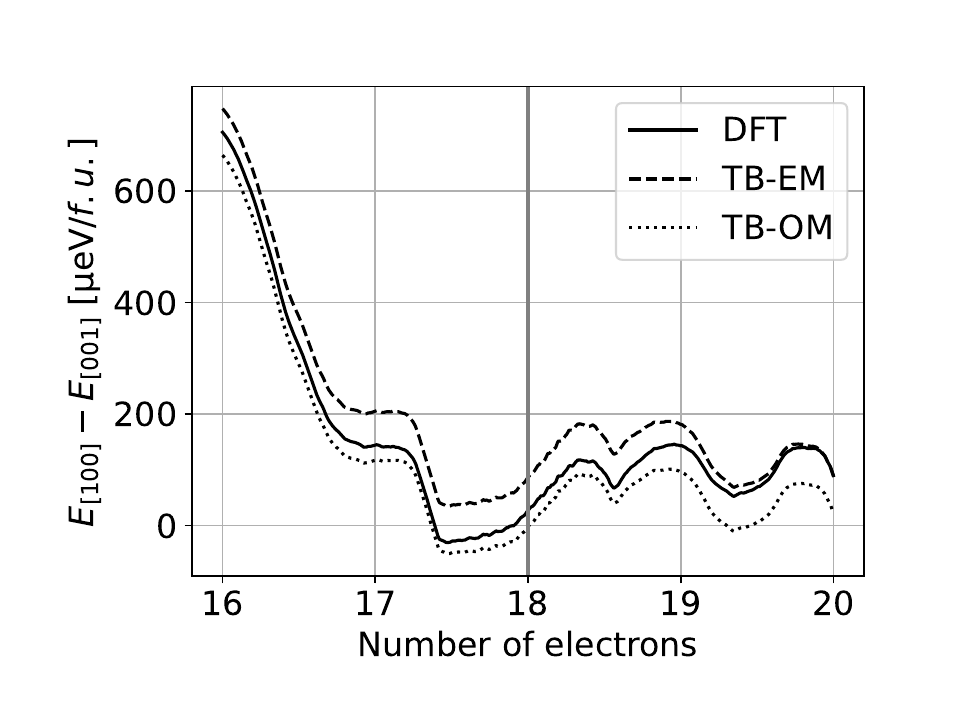}
    \end{minipage}
    \caption{Chemical potential dependence of the MCAE for (a) \(\ce{FePt}\) and (b) \(\ce{FeNi}\) calculated using the three different methods. 
        The horizontal axis represents the number of valence electrons, with the number 18 corresponding to the Fermi level.
        The DFT results (solid) are derived from Wannier \(k\) interpolation oriented in two different directions.
        In all cases, the results with $100^3$ $k$-points are shown.}
    \label{fig:bandfilling}
\end{figure}

We next focus on the difference between the TB-EM method (Method 1) and TB-OM method (Method 2), 
by calculating the chemical potential dependence of the MCAE and using the DFT method (Method 3) as a reference.
The results, presented in Fig.\ \ref{fig:bandfilling}, show that for \(\ce{FePt}\), 
the TB-EM method closely matches the DFT result within the range of valence electron numbers from 16 to 20, when compared to the TB-OM method. 
The average difference between the TB-EM and DFT results is in the order of \(10~[\mathrm{\mu eV}/f.u.]\). 
In contrast, for FeNi, both the TB-EM and TB-OM method reproduce a general trend of the DFT result, 
and the average differences with respect to the DFT results are in the order of \(10~[\mathrm{\mu eV}/f.u.]\). 
This indicates that the limit of accuracy for MCAE calculations using the TB-EM method is in the order of \(10~[\mathrm{\mu eV}/f.u.]\). 

\subsection{SOI dependence}
\begin{figure}[H]
    \captionsetup[sub]{justification=justified,singlelinecheck=false} % subcaptionの位置を左に設定
    \begin{minipage}[b]{0.48\columnwidth}
    \centering
    \subcaption{}
    \includegraphics[scale=0.43]
    {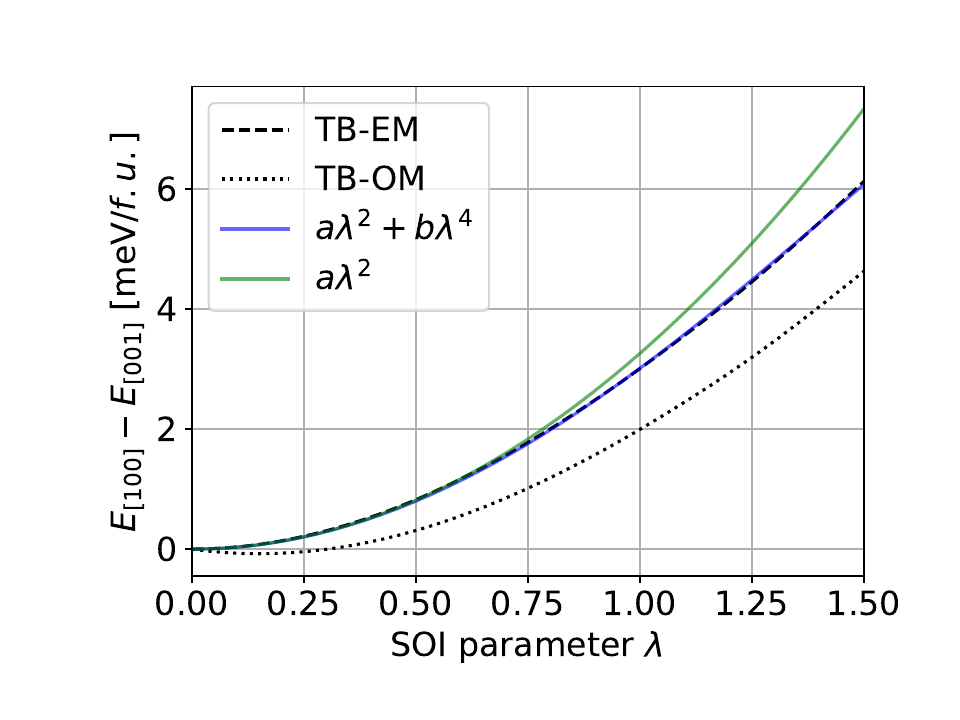}
    \end{minipage}
    \begin{minipage}[b]{0.48\columnwidth}
    \centering
    \subcaption{}
    \includegraphics[scale=0.43]
    {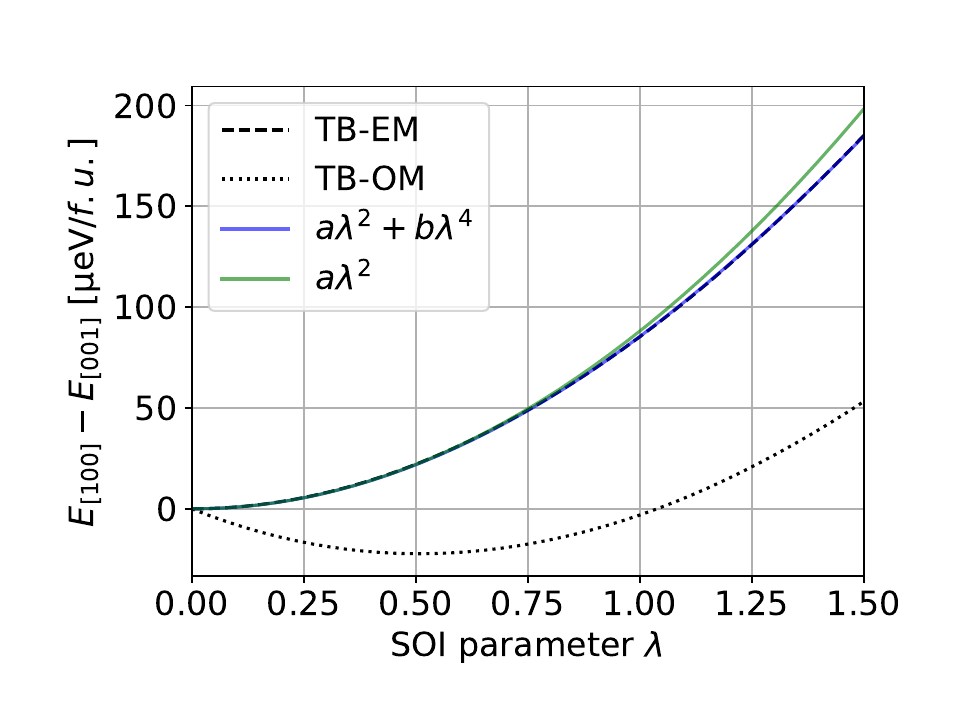}
    \end{minipage}
    \caption{Dependence of the MCAE on the magnitude of SOI, denoted as \(\lambda\), for (a) $\ce{FePt}$ and for (b) $\ce{FeNi}$. 
    Here, \(\lambda=0\) corresponds to the non-relativistic limit, while \(\lambda=1\) represents the value in actual materials. 
    The blue lines indicate the results of fitting with a fourth-order polynomial, and the green plots correspond to the situation when \(b=0\).
    The fitting parameters of the fourth-order polynomial are \(a=3.264 \pm 0.012 ~[\mathrm{meV}/f.u.]\), \(b=-0.249 \pm 0.007 ~[\mathrm{meV}/f.u.]\) for $\ce{FePt}$ and
    \(a=88.73 \pm 0.11 ~[\mathrm{\mu eV}/f.u.]\), \(b=-2.83 \pm 0.07 ~[\mathrm{\mu eV}/f.u.]\) for $\ce{FeNi}$, respectively.
    In all cases, the results with $100^3$ $k$-points are shown.}
    \label{fig:lam dep}
\end{figure}

Next, to conduct a more detailed comparison between the TB-EM and the TB-OM methods, 
we modify the magnitude of SOI in the second term of the model Hamiltonian in Eq.\ (\ref{eq:ham sym}). 
This is done by uniformly scaling all elements of the SOI term with a real parameter \(\lambda\).
The results are illustrated in Fig.\ \ref{fig:lam dep}. 
In both \(\ce{FePt}\) and \(\ce{FeNi}\) cases, 
the MCAE obtained by the TB-EM method remains positive across all values of $\lambda$ in the range $0 < \lambda < 1.5$. 
In contrast, for the TB-OM method, there are regions where the MCAE becomes negative.
Furthermore, in the TB-EM method, the MCAE exhibits a parabolic behavior in regions of smaller \(\lambda\), 
aligning with the results of the second-order perturbation theory by Bruno \cite{Bruno1989-lc} and van der Laan \cite{Van_der_Laan1998-rm}. 
Additionally, the presence of a \(\lambda^4\) term in regions of larger \(\lambda\) suggests that the TB-EM method captures higher-order perturbation effects of SOI.

%% file: result1.tex
%! Author = Hiroto Saito
%! Date = 2023/12/22

\subsection{Magnetization angle dependence}
\begin{figure}[H]
    \captionsetup[sub]{justification=justified,singlelinecheck=false} % subcaptionの位置を左に設定
    \begin{minipage}[b]{0.45\columnwidth}
    \centering
    \subcaption{}
    \includegraphics[scale=0.43]
    {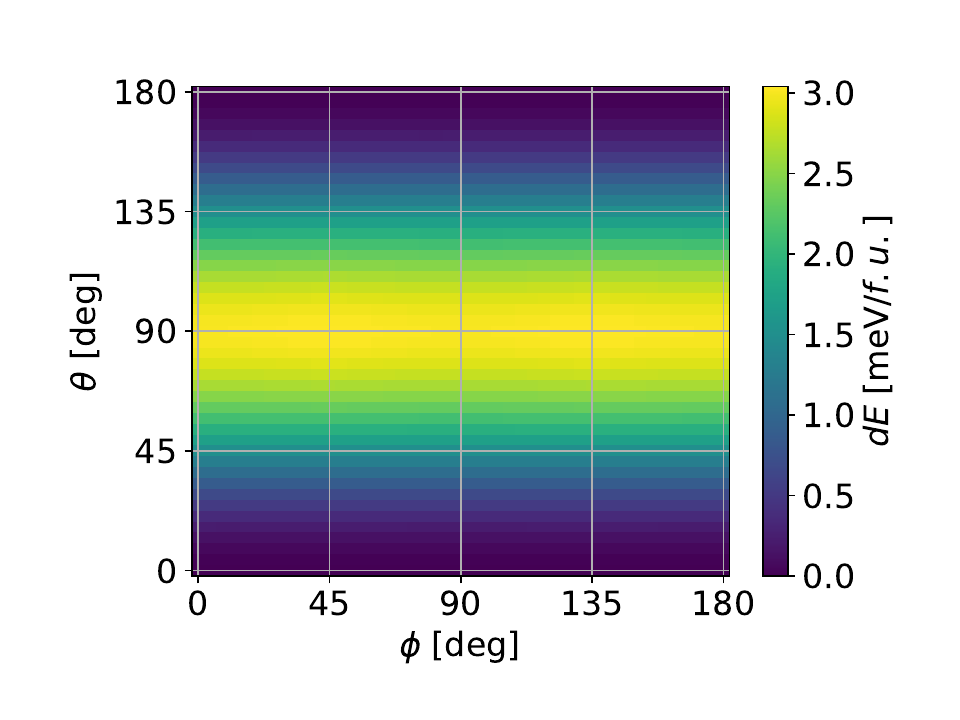}
    \end{minipage}
    \begin{minipage}[b]{0.45\columnwidth}
    \subcaption{}
    \includegraphics[scale=0.43]
    {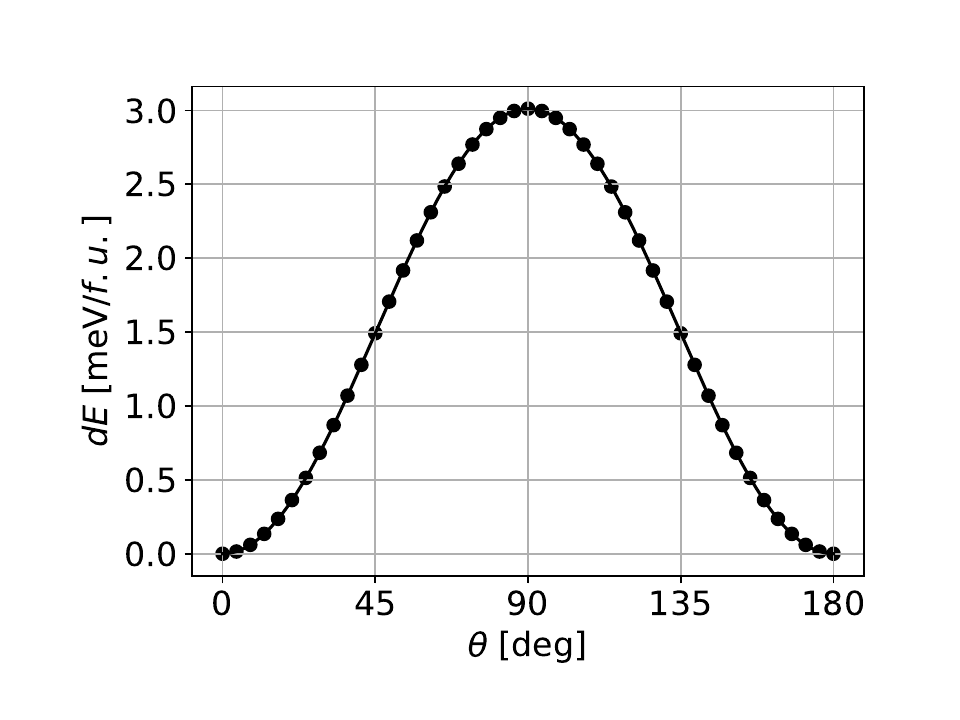}
    \end{minipage}
    \begin{minipage}[b]{0.45\columnwidth}
        \subcaption{}
    \includegraphics[scale=0.43]
    {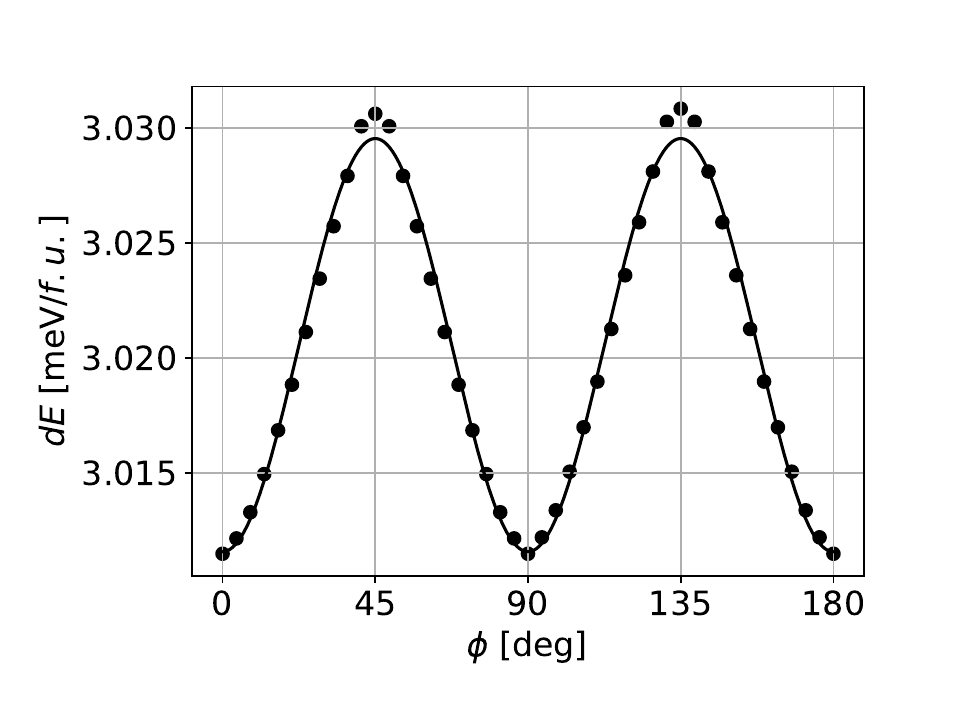}
    \end{minipage}
    \caption{Dependence of the MCAE for \(\ce{FePt}\) on the angular direction of magnetization \((\phi, \theta)\) measured from the \(c\) axis. 
    (a) For the full range of angles.
    (b) Along the line where the angle \((\phi, \theta) = (0, \theta)\).
    (c) Along the line where the angle \((\phi, \theta) = (\phi, \pi/2)\).
    In each of these cases, solid lines represent the fitting based on the phenomenological formula in Eq.\ (\ref{eq:angle dependence}). 
    The results depicted in (a) were obtained using \(60^3\) \(k\)-points, while those in (b) and (c) were derived using \(100^3\) \(k\)-points. 
    The fittings are conducted separately for (b) and (c).
    \label{fig:FePt angle dep}}
\end{figure}

\begin{figure}[H]
    \captionsetup[sub]{justification=justified,singlelinecheck=false} % subcaptionの位置を左に設定
    \begin{minipage}[b]{0.45\columnwidth}
    \centering
    \subcaption{}
    \includegraphics[scale=0.43]
    {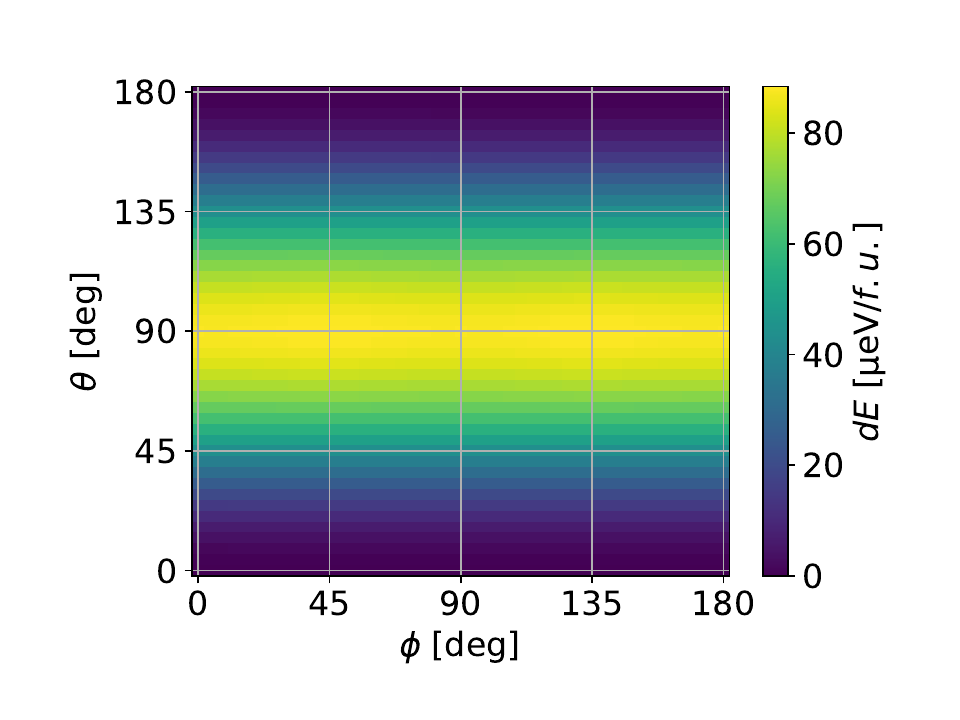}
    \end{minipage}
    \begin{minipage}[b]{0.45\columnwidth}
    \subcaption{}
    \includegraphics[scale=0.43]
    {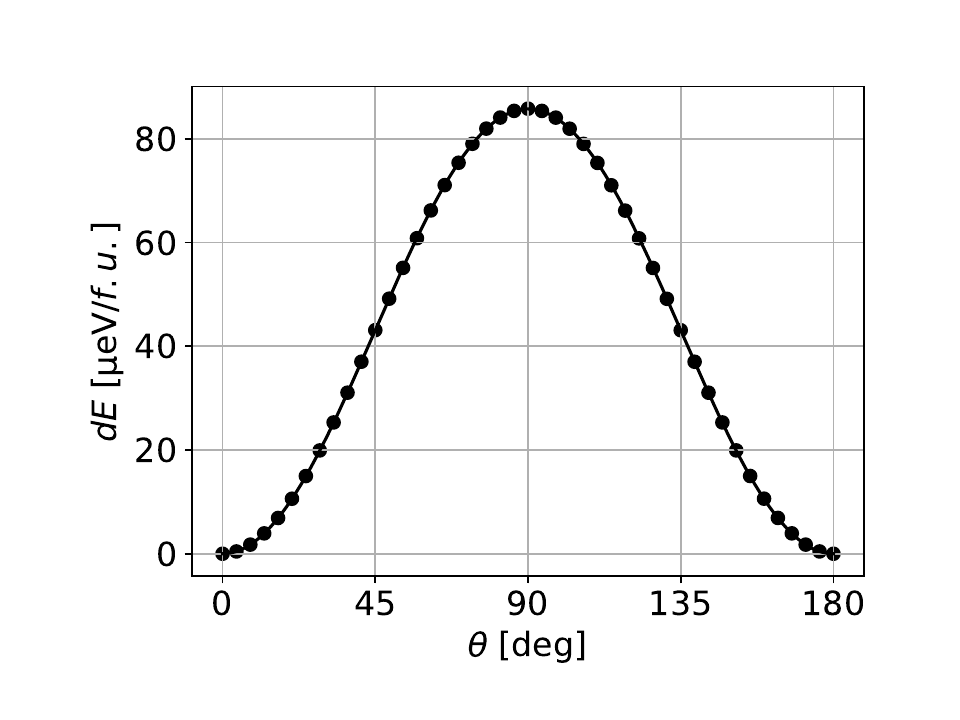}
    \end{minipage}
    \begin{minipage}[b]{0.45\columnwidth}
        \subcaption{}
    \includegraphics[scale=0.43]
    {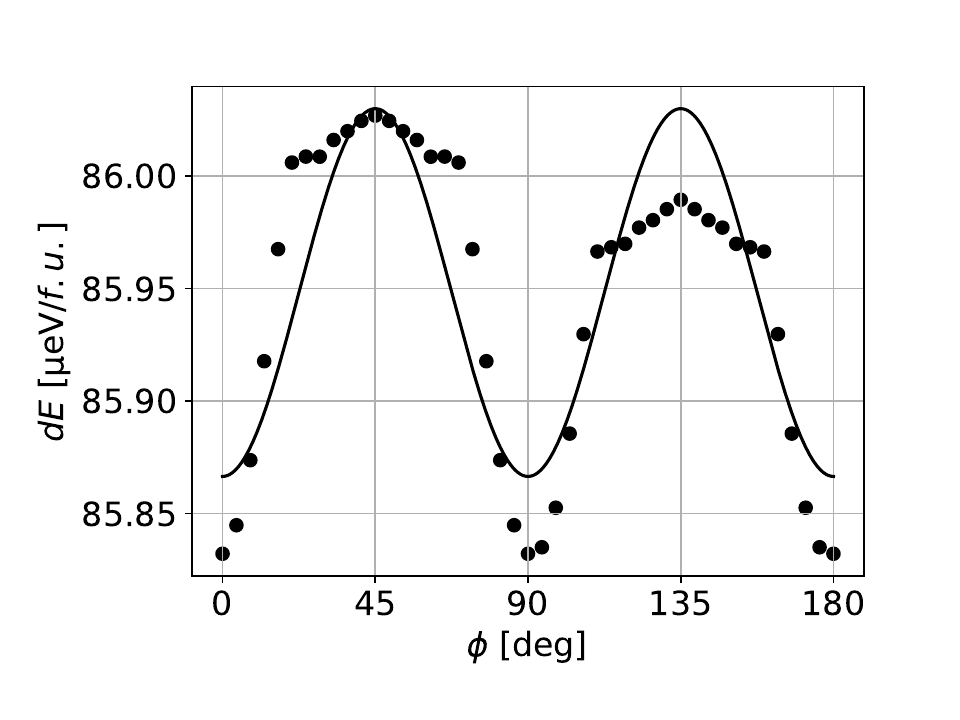}
    \end{minipage}
    \caption{Dependence of the MCAE for \(\ce{FeNi}\) on the angular direction of magnetization \((\phi, \theta)\) measured from the \(c\) axis. 
    (a) For the full range of angles.
    (b) Along the line where the angle \((\phi, \theta) = (0, \theta)\).
    (c) Along the line where the angle \((\phi, \theta) = (\phi, \pi/2)\).
    In each of these cases, solid lines represent the fitting based on the phenomenological formula, as expressed in Eq.\ (\ref{eq:angle dependence}). 
    The results depicted in (a) were obtained using \(60^3\) \(k\)-points, while those in (b) and (c) were derived using \(100^3\) \(k\)-points. 
    The fittings are conducted separately for (b) and (c).
    \label{fig:FeNi angle dep}}
\end{figure}

Finally, we calculate the detailed dependence of the MCAE on the magnetization direction \((\theta, \phi)\) using the TB-EM method. 
The angle dependence of MCAE for \(\ce{FePt}\) and \(\ce{FeNi}\) is depicted in Figs.\ \ref{fig:FePt angle dep} and \ref{fig:FeNi angle dep}, respectively.
Furthermore, we employ a phenomenological formula for the angular dependence of MCAE, 
\begin{align}
    \Delta E_{\mathrm{MCA}} (\theta,\phi)
    = K_1 \sin^2\theta + K_2 \sin^4\theta + K_3 \sin^4\theta \cos4\phi \label{eq:angle dependence}
\end{align}
to fit these data, 
from which the magnetic anisotropy constants \(K_i\) are calculated. 
The results obtained are compiled in Tabs.\ \ref{tab:FePt} and \ref{tab:FeNi}, alongside values from prior theoretical and experimental studies.

According to Fig.\ \ref{fig:FePt angle dep}, it can be seen that for \(\ce{FePt}\), the angular dependence of the MCAE aligns well with Eq.\ (\ref{eq:angle dependence}). 
The result of fitting in Fig.\ \ref{fig:FePt angle dep}(b) yields the magnetic anisotropy constants of $K_1 = 2.959~[\mathrm{meV}/f.u.]$ and $K_2 = 0.062~[\mathrm{meV}/f.u.]$, 
while that in Fig.\ \ref{fig:FePt angle dep}(c) yields $K_3 = -0.009~[\mathrm{meV}/f.u.]$.
These results, when converted into standard units $[\mathrm{MJ/m^3}]$, are compared with previous studies in Tab.\ \ref{tab:FePt}.
Notably, there are no prior studies that have investigated the \(\phi\) direction dependence given by \(K_3\), while the values and ratios of \(K_1\) and \(K_2\) show good agreement with the theoretical results of Khan \cite{Ayaz_Khan2016-qr}. 

\begin{table}[H]
    \caption{Magnetic anisotropy constants of \(\ce{FePt}\) are presented in units of \([\mathrm{MJ/m^3}]\). 
    The upper section shows theoretical values, while the lower section shows experimental results.
    \label{tab:FePt}}
    \centering
    \resizebox{\linewidth}{!}{%
    \begin{tabular}{lcccccc}
        \hline
        \hline
        Reference & $T$ [K] & $K_2/K_1$ [\%] & $K_U$ & $K_1$ & $K_2$ & $K_3$  \\
        \hline 
        this study & 0 &  2.1 & 17.964 & 17.648 & 0.370 & -0.054  \\
        \hline
        Daalderop $et~al.$ (1991) \cite{Daalderop1991-ul} & 0 & & 20.4 & & & \\ % 3.5meV
        Sakuma (1994) \cite{Sakuma1994-em} & 0 & & 16.3 & & & \\ % 2.8meV
        Solovyev $et~al.$ (1995) \cite{Solovyev1995-lt} & 0 & & 19.6 & & & \\ % 3.37meV
        Oppeneer (1998) \cite{Oppeneer1998-fb} & 0 & & 15.9 & & & \\ % 2.75meV
        Galanakis $et~al.$ (2000) \cite{Galanakis2000-jm} & 0 & & 22.1 & & & \\ % 3.90meV
        Ravindran $et~al.$ (2001) \cite{Ravindran2001-ts} & 0 & & 15.52 & & & \\ % 2.734meV
        Staunton $et~al.$ (2004) \cite{Staunton2004-wn} & 0 & & 9.523 & & & \\ % 1.696meV
        Burkert $et~al.$ (2005) \cite{Burkert2005-lu} & 0 & & 16.12 & & & \\ % 2.84meV
        Lu $et~al.$ (2010) \cite{Lu2010-ee} & 0 & & 16.62 & & & \\ % 2.90meV
        Kosugi $et~al.$ (2014) \cite{Kosugi2014-zl} & 0 & & 18.25 & & & \\ % 3.145meV
        Khan $et~al.$ (2016) \cite{Ayaz_Khan2016-qr} & 0 & & 16.59 (Wien2k) & & & \\ % 2.85meV
        Khan $et~al.$ (2016) \cite{Ayaz_Khan2016-qr} & 0 & 3.1 & 18.17 (SPR-KKR) & 17.51 & 0.54 & \\ % 3.12meV
        Ke (2019) \cite{Ke2019-lm} & 0 & & 14.51 & & & \\ % 2.495meV
        Alsaad $et~al.$ (2020) \cite{Alsaad2020-jf} & 0 & & 7.6 & & & \\ % 2.66meV
        % Nguyen $et~al.$ (2022) \cite{Nguyen2022-yw}  & 10 $\sim$ 50 & few $\sim$ 20  & & & & \\ 
        \hline 
        % Ivanov (1973) \cite{Ivanov1973-ms} & & & 7.753 & & & \\ %1.3meV
        Okamoto $et~al.$ (2002) \cite{Okamoto2002-np} &Room T. & 31.5 & & 2.13 & 0.67 & \\
        Inoue $et~al.$ (2006) \cite{Inoue2006-cp} & 5 & 1.8 & 6.9 & 7.4 & 0.13 & \\
        % Imada $et~al.$ (2007) &  &  &  & & & \\
        Richter $et~al.$ (2011) \cite{Richter2011-uo} & 850 & 9 & & 2.93 & & \\
        Goll $et~al.$ (2013) \cite{Goll2013-tu} & Room T. & & 6 & & \\
        Ikeda $et~al.$ (2017) \cite{Ikeda2017-ky} & Room T. & & 4.6 & & & \\
        Ono $et~al.$ (2018) \cite{Ono2018-pp} & Room T. & 44 & 3.4 & 0.9 & 0.4 & \\
        Saito $et~al.$ (2021) \cite{Saito2021-jh} & Room T. & & $1.8 \sim 2.3$ & & & \\
        % Tamion $et~al.$ (2022)  & & & & & \\
        \hline
        \hline
        \end{tabular}
        }
\end{table}

In contrast to the case of \(\ce{FePt}\), the angular dependence of the MCAE for \(\ce{FeNi}\),
especially in the \(\phi\) direction on the order of \(0.1~[\mathrm{\mu eV}/f.u.]\), is obscure by computational error as shown in Fig.\ \ref{fig:FeNi angle dep}.
While the positions of the peaks are accurately captured, the curve does not replicate the expected cosine curve. 
The magnetic anisotropy constants fitted using Eq.\ (\ref{eq:angle dependence}) as shown in Figs.\ \ref{fig:FeNi angle dep}(b) and (c) are found to be 
\(K_1 = 86.681~[\mathrm{\mu eV}/f.u.]\), \(K_2 = -0.772~[\mathrm{\mu eV}/f.u.]\), and \(K_3 = -0.082~[\mathrm{\mu eV}/f.u.]\).
As shown in Tab.\ \ref{tab:FeNi}, the previous study of Woodgate \cite{Woodgate2023} concluded that \(K_3\) is positive. 
This suggests that the easy axis direction in the $a$-$b$ plane for \(\ce{FeNi}\) is along \([110]\) rather than \([100]\). 
However, our calculations indicate that, similar to \(\ce{FePt}\), the easy axis in the $a$-$b$ plane for \(\ce{FeNi}\) is also in the \([100]\) (or \([010]\)) direction. 

\begin{table}[H]
    \caption{Magnetic anisotropy constants of \(\ce{FeNi}\) are presented in units of \([\mathrm{MJ/m^3}]\). 
    The upper section shows theoretical values, while the lower section shows experimental results.
    The table includes an excerpt from the sources \cite{Qiao2019-nc, Woodgate2023}.
    \label{tab:FeNi}}
    \centering
    \resizebox{\linewidth}{!}{%
    \begin{tabular}{lccccc}
        \hline
        \hline
        Reference & $T$ [K] & $K_U$ & $K_1$ & $K_2$ & $K_3$ \\
        \hline 
        this study & 0 & 0.61 & 0.6172 & -0.0055 & -0.0006 \\
        \hline
        Ravindran $et~al.$ (2001) \cite{Ravindran2001-ts} & 0 & 0.54  & & & \\ 
        Miura $et~al.$ (2013) \cite{Miura2013-nl} & 0 & 0.56  & & & \\
        Edstr\"om $et~al.$ (2014) \cite{Edstrom2014-ok} & 0 & 0.48 (Wien2k) & & & \\
        Edstr\"om $et~al.$ (2014) \cite{Edstrom2014-ok} & 0 & 0.77 (SPR-KKR) & & & \\
        Qiao $et~al.$ (2019) \cite{Qiao2019-nc} & 0 & $0.52 \pm 0.05$ & & & \\
        Ke (2019) \cite{Ke2019-lm} & 0 & 0.47 & & & \\
        Woodgate $et~al.$ (2023) \cite{Woodgate2023} & 0 & 0.96 & 0.9582 & -0.0008 & 0.0001 \\
        \hline
        N\'eel $et~al.$ (1964) \cite{Neel1964-el} & 293 & 0.55 & 0.32 & 0.23 & \\
        Paulev\'e $et~al.$ (1968) \cite{Pauleve1968-qo} & Room T. & 0.55 & 0.3 & 0.17 & 0.08 \\
        Shima $et~al.$ (2007) \cite{Shima2007-fm} & Room T. & 0.63 & & & \\
        Mizuguchi $et~al.$ (2011) \cite{Mizuguchi2011-ea} & Room T. & 0.58 & & & \\
        Kojima $et~al.$ (2011) \cite{Kojima2011-vi} & Room T. & $0.5 \pm 0.1$ & & & \\
        Frisk $et~al.$ (2017) \cite{Frisk2017-az} & 300 & $\sim 0.35$ & & & \\
        Ito $et~al.$ (2023) \cite{Ito2023-gr} & Room T. & 0.55 & & & \\
        Nishio $et~al.$ (2023) \cite{Nishio2024-di} & 300 & 0.63 & & & \\
        \hline
        \hline
    \end{tabular}
    }
\end{table}

%% file: conclusion.tex
%! Author = Hiroto Saito
%! Date = 2023/12/22

\section{Conclusion}
In our research, we developed the TB-EM method, based on the SAWFs. 
By investigating the dependence of the MCAE on the number of \(k\) points, chemical potential and SOI, 
we demonstrated that the TB-EM method provides a valid effective Hamiltonian with an accuracy of the order of approximately \(10 ~[\mathrm{\mu eV}/f.u.]\). 
Furthermore, using the TB-EM method, we determined the magnetic anisotropy constants for \(\ce{FePt}\) and \(\ce{FeNi}\).
This methodology holds the potential for low-cost computation of various physical quantities dependent on the direction of magnetization.

\begin{comment}
\section{Author declaration}
\subsection{Conflict of interest}
The authors have no conflicts to disclose.

\subsection{Author contributions}
Hiroto Saito: Methodology, Software, Validation, Formal analysis, Investigation, 
Data Curation, Writing - Original Draft, Visualization.

Takashi Koretsune: Conceptualization, Software, Resources, Writing - Review and Editing, 
Supervision, Project administration, Funding acquisition.

\subsection{Data availability}
Data will be made available on request.
\end{comment}

\section{Acknowledgement}
This work was supported by 
JSPS KAKENHI Grant No. 21H01003, 21H04437, 22K03447, and 23H04869, 
JST-Mirai Program (JPMJMI20A1), 
Center for Science and Innovation in Spintronics (CSIS), 
Tohoku University and GP-Spin at Tohoku University.

%% file: appendix.tex
%! Author = Hiroto Saito
%! Date = 2023/12/22

\appendix

\section{Band structure and density of states}
\label{app:band and dos}
\begin{figure}[H]
    \captionsetup[sub]{justification=justified,singlelinecheck=false} % subcaptionの位置を左に設定
    \begin{minipage}[b]{0.48\columnwidth}
    \centering
    \subcaption{}
    \includegraphics[scale=0.43]
    {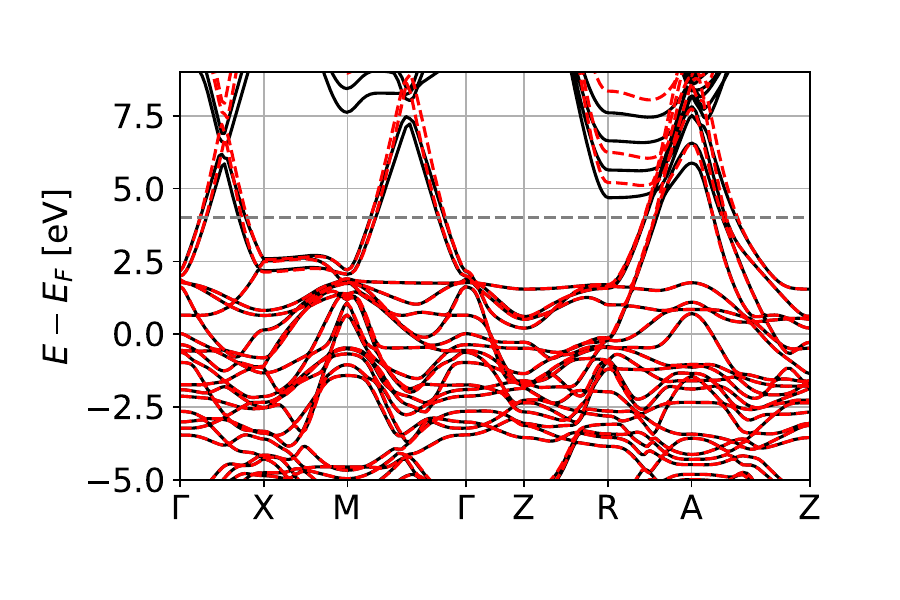}
    \end{minipage}
    \begin{minipage}[b]{0.48\columnwidth}
    \subcaption{}
    \includegraphics[scale=0.43]
    {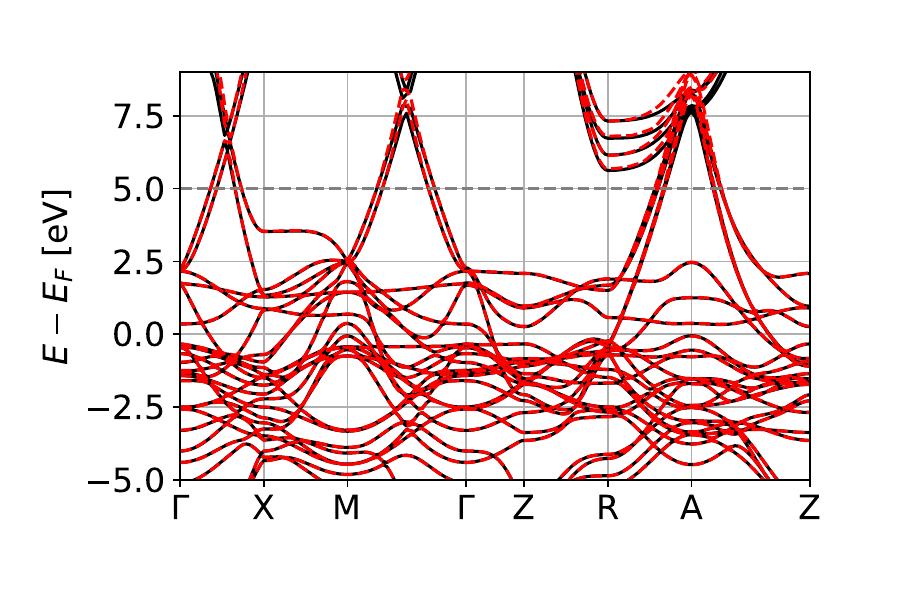}
    \end{minipage}
    \caption{
    Comparison between DFT band structure (black) and Wannier interpolated band structure (red) for (a) \(\ce{FePt}\) and (b) (a) \(\ce{FeNi}\).
    In both cases, dashed lines represent the maximum values of the inner window. 
    \label{fig:band}}
\end{figure}

\begin{figure}[H]
    \captionsetup[sub]{justification=justified,singlelinecheck=false} % subcaptionの位置を左に設定
    \begin{minipage}[b]{0.48\columnwidth}
    \centering
    \subcaption{}
    \includegraphics[scale=0.43]
    {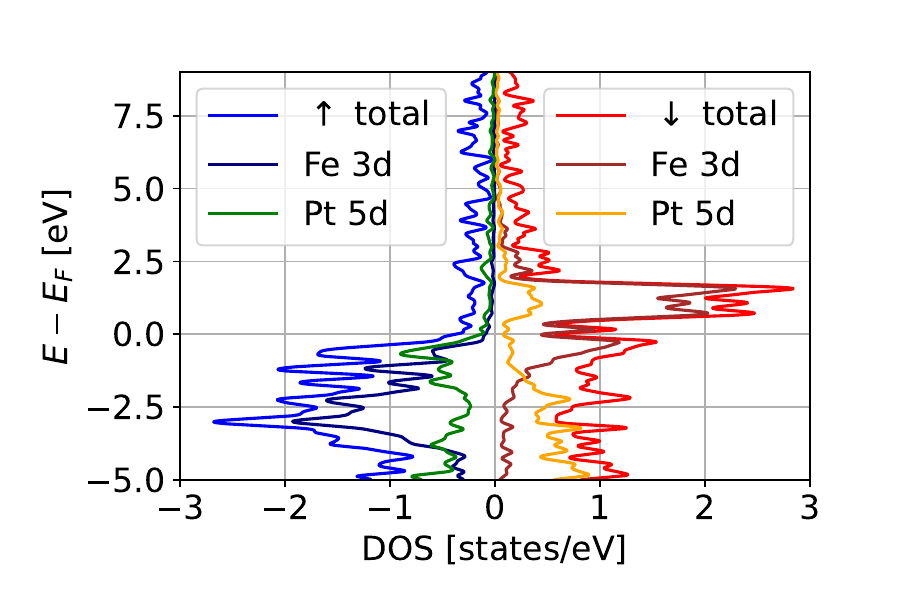}
    \end{minipage}
    \begin{minipage}[b]{0.48\columnwidth}
    \subcaption{}
    \includegraphics[scale=0.43]
    {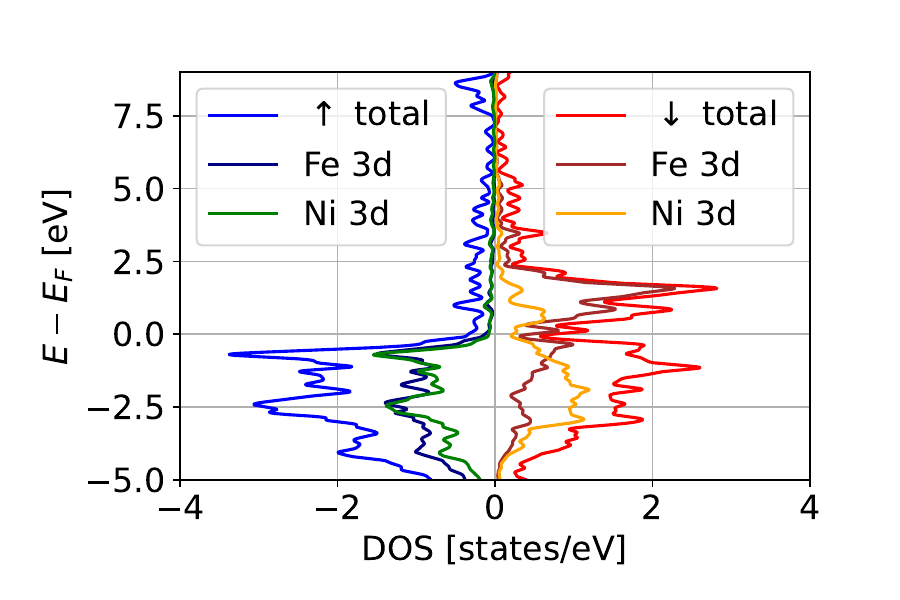}
    \end{minipage}
    \caption{Density of states for
    (a) $\ce{FePt}$ and 
    (b) $\ce{FeNi}$.}
    \label{fig:dos}
\end{figure}

Figures \ref{fig:band} and \ref{fig:dos} display the band structures and density of states for \(\ce{FePt}\) and \(\ce{FeNi}\), respectively. 
As mentioned in section \ref{sec:calculation details}, the inner windows for Wannierization were \([-1000, E_F + 4] ~[\si{eV}]\) and \([-1000, E_F + 5] ~[\si{eV}]\) for each case. 
These ranges effectively encompass the states of the \(\ce{Fe}\) $3d$ for both up and down spin.

